\documentclass[a4paper]{ESASPCS13Style}
\usepackage{epsfig}

\newcommand{\cmcc}{cm$^{-3}$}

\newcommand{\ergps}{erg~s$^{-1}$}

\newcommand{\xmm}{\textit{XMM-Newton}}

\begin{document}

\title{A Multiwavelength Study of RZ Cassiopeiae: The XMM-Newton/VLA Campaign}
\author{M.~Audard\inst{1} \and J.~R.~Donisan\inst{1} \and
M. G\"udel\inst{2}}

  \institute{Columbia Astrophysics Laboratory, Columbia University, Mail code 5247, 550 West 120th
  Street, New York, NY 10027, USA \and
  Paul Scherrer Institut, Villigen \& W\"urenlingen, 5232 Villigen PSI,
  Switzerland}

\maketitle 

\begin{abstract}

\xmm\   and the VLA simultaneously observed the eclipsing Algol-type binary 
\object{RZ Cassiopeiae} in August 2003. The secondary eclipse (K3 IV companion behind 
the A3 V primary) was placed at the center of the 15-hour radio campaign, while 
the X-ray satellite monitored a full 1.2-day orbital period. We present 
 results of the X-ray and radio campaigns. The X-ray light 
curve shows significant modulation probably related to rotational modulation and
active region evolution,
and even small flares. However, the X-ray eclipse is not deep, implying that 
the coronal X-ray emitting material is spatially extended. The Reflection 
Grating Spectrometer (RGS) spectrum shows a variety of bright emission lines from Fe, 
Ne, O, N. A strong [C/N] depletion probably reflects the surface composition of the secondary which 
fills its Roche lobe and loses material onto the primary. The O~\textsc{vii} He-like 
triplet reflects a low forbidden-to-intercombination ratio; while it generally 
suggests high electron densities, the ratio is here modified by photoexcitation 
by the strong UV flux of the primary A3 V star. The radio light curve shows no similarity
to the X-ray light curve. The eclipse timings are different, and the radio flux increased
while the X-ray flux decreased. The radio spectral slope is shallow ($\alpha = 0 - 1$).

\keywords{Stars: activity -- Stars: coronae -- Stars: individual: \object{RZ
Cas} -- radio: stars -- X-rays: stars}

\end{abstract}

\section{Introduction}
  
Direct imaging of stars still remains exceptional, and approximate geometric 
information on the coronae of stars must be obtained from indirect methods 
such as image reconstruction from rotational modulation or eclipse mapping.   
The latter is a powerful means to image the corona of a star when 
applied to Algol-type systems which are semi-detached binaries composed of
a massive early-type star orbiting an evolved late-type star. The
latter fills its Roche lobe, hence a gas stream flows from the L1 Lagrange
point (\cite{hall89}). The coronal emission originates exclusively from the 
late-type companion; therefore, the simultaneous X-ray and radio eclipses at phase 0.5 
can constrain the location of the coronal emitting regions and provide powerful 
information on the co-spatiality of the  emission regions, with important
implications for the structure  of the outer atmospheres.

In Algol-type systems, the formation of an accretion disk around the 
primary depends on the location of the binary in the $r$-$q$ diagram, where 
$r$ is the radius of the primary in units of the binary separation, and $q$ is 
the mass ratio of the secondary to the primary (\cite{peters89}). In brief, 
short-period Algols, such as our target \object{RZ Cas}, tend to populate a region of 
the $r$-$q$ diagram where the gas stream directly impacts on the primary while 
eventually forming a transient accretion disk. Although the presence of circumstellar material 
has been indirectly inferred from H$\alpha$ difference profiles (\cite{richards99}), 
its detection is still elusive in the X-rays. However, cool 
material can absorb X-rays if located along the line of sight, thus producing an 
extra column density detectable in X-rays.
  
We present results of a multiwavelength coordinated campaign
on \object{RZ Cas} in X-rays with \xmm\  and in radio
with the VLA.

\section{RZ Cassiopeiae}
 RZ Cas is a nearby ($d = 63.54$~pc) Algol-type binary, 
consisting of an A3~V primary and a K3~IV secondary with similar radii 
($R_1=1.67~R_\odot$, $R_2=1.94~R_\odot$, $A=6.77~R_\odot$; \cite{maxted94}), 
orbiting with synchronous rotation ($P \approx 1.195$~d; \cite{narusawa94}). 
Their masses are, however, different, with $M_1 = 2.2~M_\odot$ and $M_2 = 0.73~M_\odot$.
The secondary fills its Roche lobe, and a flow of material 
falls directly onto the primary (\cite{olson82}; \cite{richards99}). 
The orbital inclination angle ($i = 83.3^{\circ}$) leads to almost complete 
eclipses.

\cite*{singh95} found that RZ Cas displays variable X-ray
luminosity, with a spread of an order of magnitude ($\log L_X = 30.36 - 31.18$~\ergps).
They also found that a two-temperature optically thin plasma model
was required to fit their {\it ASCA} and {\it ROSAT} PSPC data, along with
abundances about 0.2 times the solar photospheric values. 
Their {\it ROSAT} observations showed significant intensity variations,
apparently unrelated to the eclipse, but probably originating from either 
rotational modulation, flare activity, or inhomogeneous distributions of 
coronal structures on the secondary. A core-halo structure was proposed from radio data
by \cite*{umana99}, whereas \cite*{gunn03} observed a radio modulation
close to the primary eclipse seemingly correlating with X-rays.

\section{Observations and Data Reduction}

\begin{figure*}[!th]
\centering
\resizebox{\hsize}{!}{\includegraphics[angle=0]{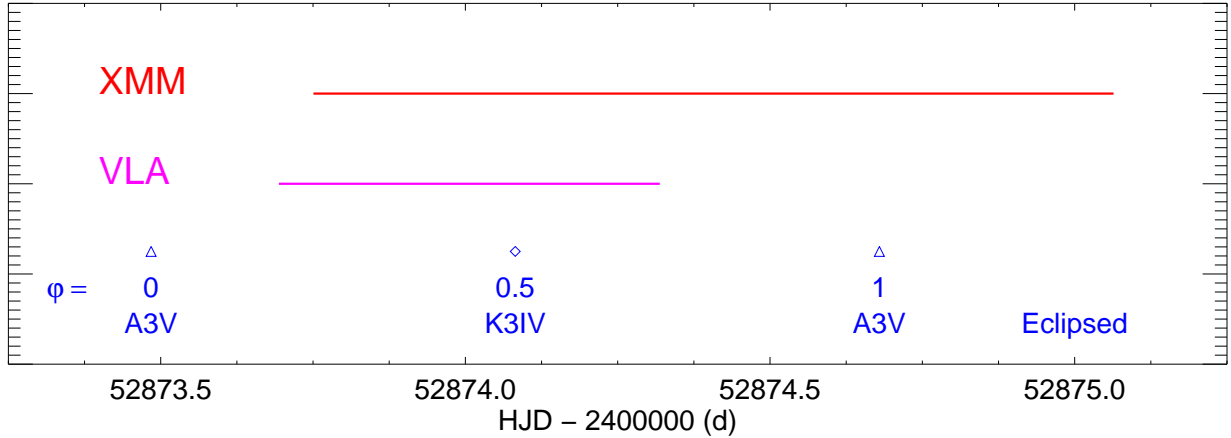}}
\caption{Schematic diagram of the coordinated campaign with \xmm\  and the VLA. The
horizontal lines show the observation time ranges for each observatory ($30.55$~hrs with
\xmm, $15$~hrs with VLA). The orbital phases are shown in the lower part, with the
eclipsed star indicated.%
\label{fig:campaign}}
\end{figure*}

\xmm\  observed RZ Cas for 110~ks from August 22 (5h23m54s UT) to August 23,
2003 (13h41m23s UT), whereas the VLA observed on August 22, 2003 from 4h41m20s 
UT to 19h38m30s UT . The VLA was in A-array configuration, and we
observed alternatively in the C (6.1 cm) and X (3.55 cm) bands, with 0137+331 (3C 48) as flux
calibrator and 0228+673 as phase calibrator. Figure~\ref{fig:campaign} shows a
diagram of the coordinated observations.

The X-ray data were reduced with the \xmm\  SAS software (version 6) using standard procedures. 
The radio data were reduced with AIPS (December 31, 2003 version)
using standard techniques. The phases before 9h UT and after 16h UT were
significantly degraded due to strong cumuloform clouds. No
reliable solutions could be found during these periods (see shaded regions in Fig.~\ref{fig:vlaxmm}). 
We used different techniques to obtain the radio light curves in both bands and to check their consistency.

\section{Light curves}
\begin{figure*}[!ht]
\centering
\resizebox{\hsize}{!}{\includegraphics[angle=0]{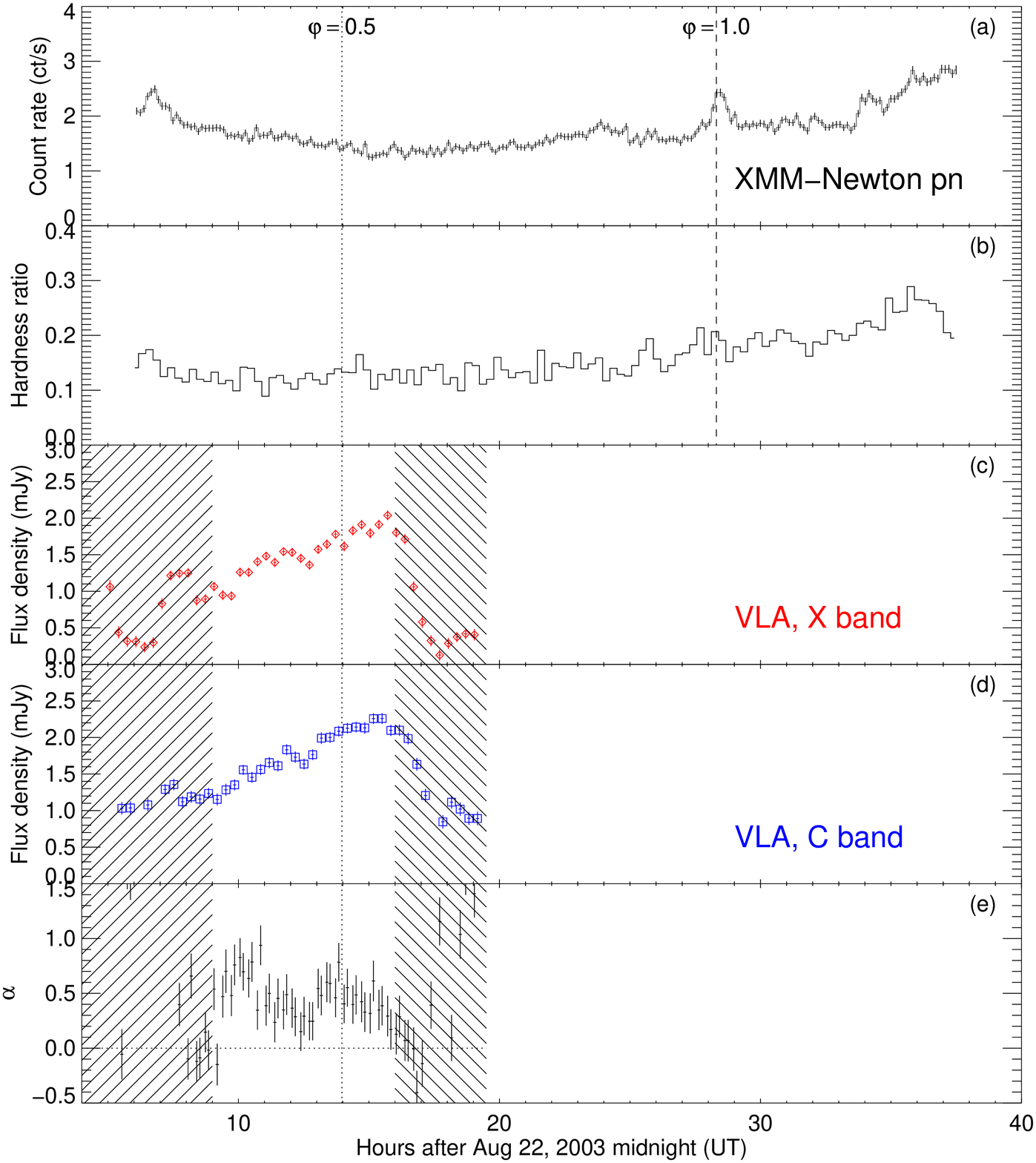}}
\caption{(a) \xmm\  pn light curve in the $0.25-8$~keV range, and (b) its hardness ratio
for a energy separation at $1.5$~keV. Although the ratio does not 
change significantly during the eclipse, it consistently increases after phase $0.5$,
which is confirmed by higher plasma temperatures in the later time ranges of the 
observation. The radio VLA light curves in the X (c) and C (d)
bands are also shown. The spectral index $\alpha$ (assuming a power-law shape $S \sim
\nu^{-\alpha}$) is shown in the  panel (e). A shallow index is observed throughout the
observation, consistently with the core-halo model. A possible 
flattening during the secondary eclipse in the radio is observed. The shallow eclipses observed
in both regimes occur, however, at different times.\label{fig:vlaxmm}}
\end{figure*}

Figure~\ref{fig:vlaxmm} shows the X-ray pn
light curve for events from 0.25 to 8~keV (panel a) and a light curve of a hardness ratio
(panel b) defined here as the ratio between the count rates in the hard band
($1.5-8$~keV) and the soft band ($0.25-1.5$~keV). The X-ray eclipse is almost absent,
with a local X-ray minimum about an hour \emph{after} the optical eclipse of the K3 subgiant.
The light curve modulation suggests, instead, the presence of extended X-ray sources
which are not significantly eclipsed and are rotationally modulated (e.g., in particular
in the phase range $\varphi = 0.25-0.75$). It could also reflect the evolution of active regions
in RZ Cas. A flare was observed at $\varphi = 0$, i.e.,
at the time of the eclipse of the primary A3~V star. However, the synchronous rotation
of the stars does not suggest that $\varphi = 0$ should be a special orbital phase for
flares. We found also that RZ Cas significantly became significantly and steadily hotter in the second half
of the observation, as shown in the hardness ratio light curve. No obvious flare 
can explain such hardening, therefore we hypothesize that hot, active regions rose in
this time span, probably due to some reconfiguration of the magnetic topology in RZ Cas.

The radio light curves in X and C bands are also shown in Figure~\ref{fig:vlaxmm} in panels (c) and (d),
respectively. A similar flux level was observed in both frequencies. Panel (e) of Fig.~\ref{fig:vlaxmm} shows
a light curve of the spectral index $\alpha$ (assuming a power law for the radio
spectrum, i.e., $S \sim \nu^{-\alpha}$). The shaded time ranges in the radio light
curves correspond to time spans when an accurate calibration could not be obtained.
The index $\alpha$ appears to be shallow
($\alpha = 0-1$) between 4.9~GHz and 8.4~GHz. This is consistent with the core-halo
model proposed by \cite*{umana99} for RZ Cas. The radio light curves show a steady
flux increase, an a possible shallow
eclipse is observed about 1.5~hours \emph{before} the optical secondary eclipse. The
spectral index $\alpha$ possibly flattens during the radio eclipse.

The marked difference in i) eclipse timing, and ii) in flux
modulation in the X-rays and radio strongly suggests that the emitting
sources are not co-spatial, at least during this coordinated campaign. 
This does not support the findings of \cite*{gunn03}. In addition, the shallow
eclipses preclude compact sources in both wavelength regimes. Furthermore, the absence of
eclipse at phase $\varphi = 0$ indicates that the primary A-type star does not emit
significant X-rays. Finally, the lack of significant eclipses at both primary and
secondary eclipses precludes intrabinary (or accretion spot) X-ray and radio
emissions in RZ Cas. Instead, the light curves imply rotational modulation of extended
sources and active regions rise in the second half of the X-ray observation.

\section{X-ray Spectra}

\begin{figure}[!t]
\centering
\resizebox{\hsize}{!}{\includegraphics[angle=0]{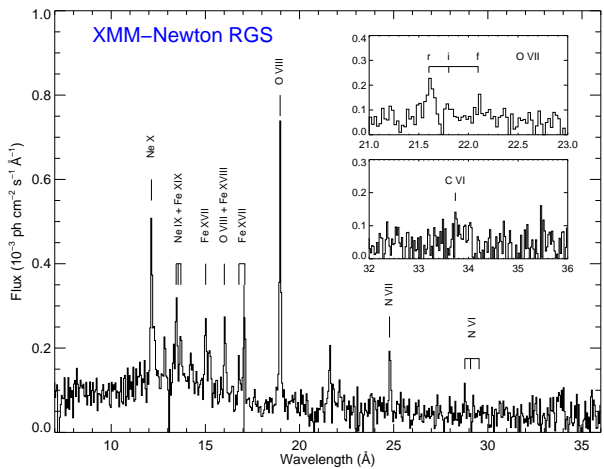}}
\caption{\xmm\  RGS fluxed spectrum of RZ Cas. Major emission lines are labeled. Notice the strong
{\rm N~\textsc{vii}} Ly$\alpha$ line and weak {\rm C~\textsc{vi}} Ly$\alpha$ line, a signature of {\rm CNO}-processed material.%
\label{fig:rgs}}
\vskip -4mm
\end{figure}

Figure~\ref{fig:rgs} shows the average \xmm\   RGS spectrum of RZ Cas.
Strong emission lines of O~\textsc{viii}, Ne~\textsc{x}, Ne~\textsc{ix}, and several
Fe L-shell lines can be observed on top of an underlying continuum. The high
Ne~\textsc{x}/Ne~\textsc{ix} and O~\textsc{viii}/O~\textsc{vii} ratios indicate a
dominant hot plasma in RZ Cas's corona. The faint O~\textsc{vii} triplet suggests a high
intercombination-to-forbidden line ratio, albeit with a large error bar. Although 
such a ratio could be indicative of a high-density ($n_\mathrm{e} = 10^9 - 10^{11}$~\cmcc) 
plasma, it is in fact artificial and due to the strong UV field from the early-type primary 
(see \cite{ness02}).
The poor signal-to-noise ratio in the triplet prevented us from obtaining a
phase-dependent line ratio, which would have confirmed the influence of the UV field
(minimal close to $\varphi = 0.5$, maximal close to $\varphi = 0$). 

We performed a multi-$T$ fit of the RZ Cas data using both MOS spectra in the
$1.55-18$~\AA\  range and both RGS spectra above $8$~\AA. We also discarded several
wavelength ranges of the RGS data to avoid bias from poorly known L-shell lines of low-Z
ions (see \cite{audard03}). A 4-$T$ model fits the data adequately ($\chi^2 = 1602$
for 1202 d.o.f.), with $T_1 = 3.6^{+0.8}_{-0.6}$~MK, $T_2 = 7.6^{+1.3}_{-5.8} $~MK, 
$T_3 = 11.6^{+14.2}_{-2.8}$~MK, $T_4 =
25.5^{+3.5}_{-1.6}$~MK, $\log \mathrm{EM_1} = 52.11^{+0.23}_{-0.25}
$ \cmcc, $\log \mathrm{EM_2} = 52.64^{+0.22}_{-1.53}$~\cmcc,
$\log \mathrm{EM_3} = 52.50^{+0.11}_{-0.55} $ \cmcc, $\log \mathrm{EM_4} = 53.03^{+0.05}_{-0.21}$~\cmcc. 
Coronal abundances (relative to the solar photospheric abundances; 
\cite{grevesse98}) are $\mathrm{C} = 0.15^{+0.15}_{-0.15}$, $\mathrm{N} =
0.85^{+0.36}_{-0.32}$, $\mathrm{O} = 0.38^{+0.06}_{-0.05}$, 
$\mathrm{Ne} = 0.97^{+0.12}_{-0.11}$, $\mathrm{Mg} = 0.68^{+0.08}_{-0.07}$, 
$\mathrm{Si} = 0.31^{+0.05}_{-0.05}$, $\mathrm{S} = 0.15^{+0.06}_{-0.06}$, 
$\mathrm{Ar} = 0.27^{+0.33}_{-0.27}$, $\mathrm{Ca} = 0.60^{+0.34}_{-0.33}$, 
$\mathrm{Fe} = 0.37^{+0.04}_{-0.04}$. All quoted uncertainties are 90 \% confidence
ranges ($\Delta\chi^2 = 2.71$). The high N/C abundance ratio is a strong signature of
CNO cycle processed material. This has been observed in Algol (\cite{drake03})
and in the rapidly rotating giant star YY Men (\cite{audard04}).

We also performed time-dependent spectroscopy using the EPIC MOS spectra only. We observed no
significant variation in abundances. However, consistently with the hardness ratio light
curve (Fig.~\ref{fig:vlaxmm}), we obtained slightly hotter plasma in the second half of
the observation, in particular in the last high-flux time segment. We also observed no
variation in the spectral slope at low energies, implying that photoelectric absorption
(e.g., due to the accretion disk near $\varphi = 0$) remained constant and
was consistent with interstellar absorption.

\begin{figure}[!t]
\centering
\resizebox{\hsize}{!}{\includegraphics[angle=-90]{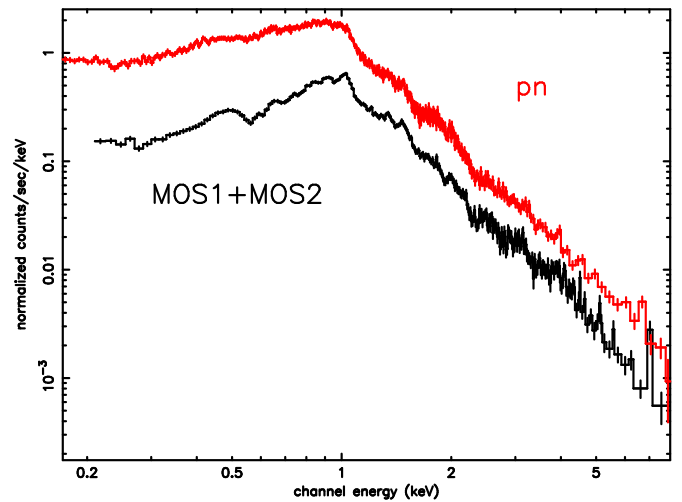}}
\caption{{\it \xmm\  EPIC average spectra of RZ Cas.}%
\label{fig:epic}}
\vskip -2mm
\end{figure}

\begin{acknowledgements}
Based on  observations obtained with \textit{XMM-Newton}, an 
ESA science mission  with instruments and contributions directly funded by 
ESA Member States and NASA. The NRAO is a facility of 
the NSF operated under cooperative agreement by Associated Universities, Inc.
M.~A. and J.~R.~D. acknowledge support from NASA grant 
NNG04GA42G, and M.~G. from the Swiss NSF (grant
20-66875.01).
\end{acknowledgements}

\end{document}